\documentclass[12pt,twoside]{article}
\usepackage{wrapfig}
\usepackage{graphicx}
\usepackage{cmp2e}
\newtheorem{defn}{Definition}
\newtheorem{thm}{Theorem}
\DeclareSymbolFont{AMSb}{U}{msb}{m}{n}
\DeclareMathSymbol{\N}{\mathbin}{AMSb}{"4E}
\DeclareMathSymbol{\Z}{\mathbin}{AMSb}{"5A}
\DeclareMathSymbol{\R}{\mathbin}{AMSb}{"52}
\DeclareMathSymbol{\Q}{\mathbin}{AMSb}{"51}
\DeclareMathSymbol{\I}{\mathbin}{AMSb}{"49}
\DeclareMathSymbol{\C}{\mathbin}{AMSb}{"43}
\title[Homotopy in statistical physics]%
{Homotopy in statistical physics%
}

  \author[R. Kenna]{R. Kenna}
  \address{Applied Mathematics Research Centre,
            Coventry University,
            Coventry, CV1 5FB, England}
\begin{document}

\maketitle

\begin{abstract}
In condensed matter physics and related areas, 
topological defects
play important roles in phase transitions and critical phenomena.
Homotopy theory facilitates the classification of such topological defects.
After a pedagogic introduction to the mathematical methods involved in topology
and homotopy theory, 
the role of the latter in a number of mainly low-dimensional statistical-mechanical systems is
outlined. Some recent activities in this area are reviewed and some 
possible future directions are discussed.
%
\pacs 02.40.Pc; 05.20.-y; 64.60.-i; 64.70.Md; 64.70.-p 
\end{abstract}
\tableofcontents

\section{Introduction}
\label{intro}

Topology is the appropriate mathematical framework for the study of spaces which can 
(and cannot) be continuously deformed into each other. Continuous deformations include 
twisting and stretching but not tearing or puncturing. Thus a cube and a sphere
are topologically equivalent entities. Similarly a square is 
equivalent to a circle in topological terms.
A square is, of course, quite different to a circle in that there is 
a lack of differentiability at its vertices. The appropriate mathematical 
framework to deal with this aspect is differential geometry.
Since differentiability necessitates continuity, differential geometry is, in a sense,
more restrictive than topology. 
However, while the former may yield more concrete results, topological arguments generally 
only lead to existence or classification statements.

This paper contains a short pedagogic review of certain topological concepts in statistical physics,
with a focus on homotopy and its consequences. The first part (sections~\ref{basic}--\ref{higherhomotopy})
summarizes the basic conceptual and calculational tools used in the determination of the homotopy groups
for simple topological spaces. The second part 
of the paper (sections~\ref{phasetransitions}--\ref{nsm}) contains a review of recent
progress in statistical mechanical models where topological concepts play a crucial role.
Conclusions are outlined in section~\ref{conc}.

\section{Basic notions of topology}
\label{basic}
We begin by introducing some basic topological notions, with the primary objective of 
being able to study continuity in mind. 
We refer the reader to the literature (e.g., \cite{NaSe88}) 
for basic proofs, which are all rather straightforward.
\begin{defn}
\label{defn1}
 If $X$ is a set and $T = \{ X_i \} $ is a collection of finitely or 
infinitely many subsets of $X$, then we say $X$ is a topological space with a
topology $T$  
%
(i) if $\emptyset \in T$, $X \in T$,
(ii) if $\{X_j \}$ is a finite or infinite subset of $T$, then the union
$\cup_j{X_j} \in T $ and
%
(iii) if $\{X_j\}$ is a finite (not infinite) subset of $T$, then the intersection
$\cap_j{X_j} \in T $.
%
The sets $X_i$ are  called open sets.
\end{defn}
With this definition, the topology consisting only of $\emptyset$ and $X$ is the 
one with the least number of open sets. 
For the largest topology, all possible 
subsets of $X$ and, indeed, all points in $X$
are open sets. The latter is called the discrete topology. 
In  Euclidean or Cartesian space $\R^n$, one more commonly employs the 
{\emph{usual}} or {\emph{ordinary}} or {\emph{Euclidean}} topology,  
in which the open sets are restricted to $n$-balls
or open intervals. Definition~\ref{defn1} is crucial when analyzing continuity,
which is the basic purpose of topology. 
The reason why infinite intersections are not allowed in the definition
is that such a construct may 
give rise to open sets under the usual topology which consist of a single point. 
This is not what we usually understand by an open interval and would render useless 
the following definition of continuity.
\begin{defn}
\label{defn2}
If $X$ and $Y$ are topological spaces and if $f: X \rightarrow Y$, then 
$f$ is continuous if the inverse of an open set in $Y$ is open in $X$.
\end{defn}
We examine this definition using an example of the usual topology in $\R$, taking $X=Y=\R$, and the
discontinuous function
\[
 f(x) = \left\{ \begin{array}{ll}
        x + 1 & \mbox{if $x \ge 0$}\,, \\
        x     &  \mbox{if $x <   0$}\,.
                \end{array}
        \right.
\]

\begin{wrapfigure}{i}{0.5\textwidth}
\vspace{-1.9cm}%
\hspace{3.0cm}
\centerline{\includegraphics[width=0.8\textwidth]{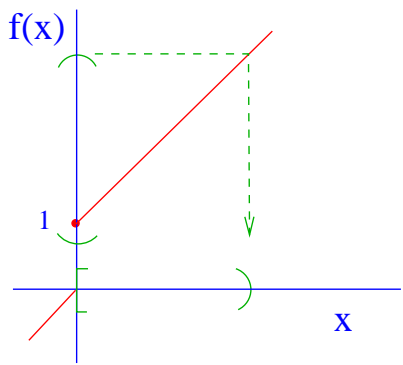}}
\vspace{-11.5cm}%
\caption{Illustration of the topological definition of continuity.}
\label{fig:continuity}
\vspace{-0.3cm}%
\end{wrapfigure}
The function is depicted in figure~\ref{fig:continuity}. It is also 
demonstrated in the figure that while the inverse of some open sets in $Y$ 
are open sets in $X$, this is not the case if the open set in $Y$ is inclusive of
part of the discontinuous zone. 
So definition~\ref{defn2} captures the notion of discontinuity.
Moreover, a moments consideration renders it clear that usage of $f$ rather than
$f^{-1}$ in definition~\ref{defn2} would be useless, as $f$ always takes open
sets (in $X$) into open sets (in $Y$). 

We now move on to a number of other basic definitions of topology. 

\begin{defn}
\label{defn3}
A neighborhood of a point $x\in X$ is a subset $N$ which contains an open set $X_i$ containing
$x$. I.e., $x \in X_i \subseteq N \subseteq X$. 
\end{defn}
Note that $N$ doesn't have to be open (for example, if $X=\R$, with the usual topology
the closed interval
$[5,7]$ is a non-open neighborhood of the point $x=6$).
But all open sets $X_i$ containing $x$ are
neighborhoods of $x$.
\begin{defn}
\label{defn4}
 A subset $U$ is closed if its complement $X-U$ is open. The complement of an open set is called closed.
\end{defn}
By definition, $X$ is open. Since its complement is $\emptyset$, the latter is closed. Similarly,
since by definition $\emptyset$ is open, its complement $X$ is closed. So $\emptyset$ and $X$
are examples of sets which are both open and closed. 
\begin{defn}
\label{defn5}
 If $U$ is a set, its closure, written $\overline{U}$, 
 is the smallest closed set in which $U$ is contained. 
 Because arbitrary intersection of closed sets results in a closed set, 
 one may write $\overline{U} = \cap_\alpha{V_i}$, where $V_i$ are all closed sets containing $U$.
\end{defn}
So, for example, the closure of the open set $(5,7)$ is the closed interval $[5,7]$.
\begin{defn}
\label{defn6}
The interior $U^0$ of a set $U$ is the union of all open subsets of $U$.
\end{defn}
Thus the interior of the closed interval $[5,7]$ is the open set $(5,7)$
and the interior of a closed disk is an open one, for example.
\begin{defn}
\label{defn7}
The boundary $b(U)$ of a set $U$ is the complement of its interior within its closure: 
$b(U) = \overline{U} - U^0$. 
\end{defn}
Note that $U \cap b(U) = \emptyset$ if $U$ is open. 
\begin{defn}
\label{defn8}
A subset $U$ of $X$ is dense in $X$ if its closure $\overline{U} =X$.
\end{defn}
For example, with the usual topology on $\R$, there are no open subsets of the set of
rationals $\Q$. So the interior $\Q^0$, which is the union of all open subsets of
$\Q$, is $\emptyset$. In this case $b(\Q) = \overline{\Q} - \Q^0 = \overline{\Q}$.
This means that the boundary is the closure. However, since $b(\Q) = \R$, we have that 
$\overline{\Q} = \R$. Therefore the set of rational numbers is dense in the reals.
\begin{defn}
\label{defn9}
Given a set $U$ and a (finite or infinite) family of sets $V=\{V_i \}$,
if $U$ is contained in $\cup_i{V_i}$ we say that $V$ is a cover of $U$. 
If all sets $V_i$ are open then $V$ is an open cover.
\end{defn}
For example, the set of open intervals $(-n,n)$, where $n \in \N$, is an open cover of 
$\R$ under the usual  topology.
\begin{defn}
\label{defn10}
A set $U$ is called compact if every open cover has a finite subcover (say 
$\{ V_1, V_2, \dots, V_N \}$), such that $U \subset V_1 \cup V_2 \cup \dots V_N$.
\end{defn}
In fact, for a set to be compact, it has to be closed and bounded (and vice versa).
Essentially this means it has finite volume.

\begin{wrapfigure}{i}{0.5\textwidth}
\vspace{-2.5cm}
\hspace{1.6cm}
\centerline{\includegraphics[width=0.7\textwidth]{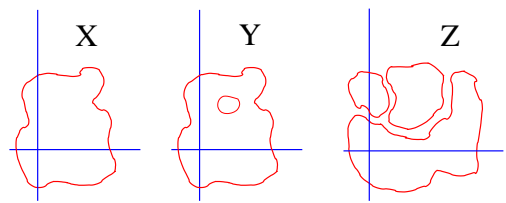}}
\vspace{-9.7cm}%
\caption{Space X is simply connected while $Y$ is connected but not simply (it is multiply 
or non-simply connected) 
and $Z$ is not conected.}
\label{fig:connected}
\end{wrapfigure}
\begin{defn}
\label{defn11}
A set $U$ is called connected if it cannot be written as 
$U = U_1 \cup U_2$ in which $U_1 \cap U_2 = \emptyset$.
It is simply connected or $1$-connected if any loop in it can be continuously 
contracted to a point. A domain is a connected open set.
\end{defn}
This definition is illustrated in figure~\ref{fig:connected}.

For a set to be simply connected  it must consist of one component 
and have no holes. Higher-dimensional holes are, however,  allowed,
such as the three-dimensional cavity enclosed by the $2$-sphere,
which is a simply connected space. 

If the set itself can be continuously contracted to a point it is said to
be contractable. All contractable sets are simply connected but the converse is not true.
For example, the sphere is simply connected but not contractable -- for, if it is reduced to a single point,
it is no longer a sphere. Contractibility
is stronger than simply connected and means the space has no holes or cavities of
any dimension.

Our central theme is the study of continuous deformations of spaces from one to another. 
In fact, much information will be gleaned from circumstances in which this is not possible.
Indeed, if such continuous deformations are prohibited, then there must be some
obstacle in the way. These are called topological invariants. 
We attempt such continuous deformations through homeomorphisms (`homeo' and `morphism'
coming from the Greek words for `similar' and `shape', respectively).
\begin{defn}
\label{defn12}
If $X$ and $Y$ are two topological spaces then $f: X \rightarrow Y$ is a 
homeomorphism if it is continuous and if there exists a
 continuous map $g: Y\rightarrow X$ such that
$f \circ g = 1_Y$ (the identity map in $Y$). 
The map $g$ is also a homeomorphism and  $g \circ f = 1_X$ (the identity in $X$).
I.e., $f = g^{-1}$ and vice versa. 
\end{defn}
\begin{wrapfigure}{i}{0.5\textwidth}
\vspace{-2.2cm}
\hspace{2.8cm}
\centerline{\includegraphics[width=0.75\textwidth]{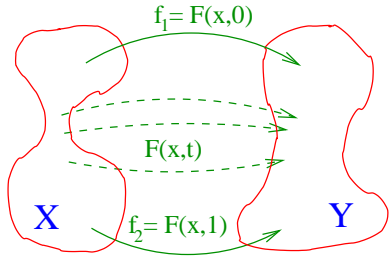}}
\vspace{-10.5cm}%
\caption{Representation of a homotopy $F(x,t)$ between two maps
$f_1$ and $f_2$.}
\label{fig:homotopy}
\vspace{-0.0cm}%
\end{wrapfigure}
\vspace{-0.0cm}%

With this concept we can categorise all topological spaces into equivalence classes. 
Spaces belong to the same class if they are homeomorphic to each other.
To characterize homeomorphic equivalence classes we need topological invariants
(which are not broken under homeomorphism). These include the dimension of the space,
properties such as compactness and connectedness and the powerful concept of homotopy,
to which we now turn (`homo' and `topos' come from the Greek words for `same' and `place').
Homotopy is to continuous maps what homeomorphism is to topological spaces
-- homotopy continuously distorts maps while homeomorphism continuously distorts spaces.
The concept is illustrated schematically in figure~\ref{fig:homotopy} in the context 
of the following definition.
\begin{defn}
\label{defn13}
Let $X$ and $Y$ be topological spaces and $f_1$ and $f_2$ be continuous
maps from $X$ to $Y$. Then $f_1$ is homotopic to $f_2$ and vice versa if
$f_1$ can be continuously deformed into $f_2$ in the sense that there exists 
a continuous function $F: X \times [0,1] \rightarrow Y$, such that 
\[
 F(x,0) = f_1(x) \;, \quad \quad \quad  F(x,1) = f_2(x) \;.
\]
\end{defn}
Homotopy is an equivalence relation and categorizes all continuous maps from $X$ to $Y$ 
into homotopy equivalence classes which are unchanged under homeomorphism. 

The various so-called homotopy groups provide deep insights into topological phenomena in 
physics, and we now introduce these. 

\section{The fundamental or first homotopy group}
\label{fund}

If spaces can be continuously deformed into each other without breaking or tearing
then they belong to the same homeomorphic equivalence class. 
Clearly, a simply connected compact space $X$ (one with no holes) is not in
the same homeo\-morphic equivalence class as a non-simply connected space $Y$ (which has a hole).
To quantify this rather intuitive statement we consider loops and classes of loops within 
each space.

\begin{wrapfigure}{i}{0.5\textwidth}
\vspace{-3.2cm}
\hspace{3.0cm}
\centerline{\includegraphics[width=0.85\textwidth]{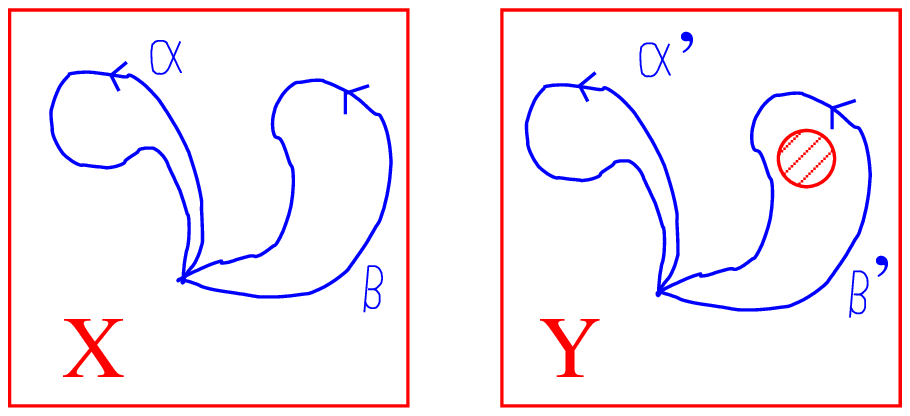}}
\vspace{-12.6cm}%
\caption{All loops in space $X$ are homotopic. This is not the case in space $Y$, which has a hole.}
\label{Fig:holes}
\end{wrapfigure}

Referring to figure~\ref{Fig:holes}, it is clear that while loops $\alpha$ and $\beta$
in space $X$ and $\alpha^\prime$ in space $Y$ can be shrunk to a point, $\beta^\prime$ in $Y$ cannot
(because the hole encompassed by  $\beta^\prime$ forms an obstruction).
We say $\alpha$ and $\beta$ are homotopic while $\alpha^\prime$ and $\beta^\prime$ are not, 
and write
\[
 \alpha \simeq \beta \;, \quad 
 \alpha^\prime  \simeq \!\!\!\!\! / \,\,\, \beta^\prime
\;.
\]
In fact every loop  in space $X$ can be continuously shrunk to a single point 
(the constant or identity loop) and
there is one homotopy class of loops there. In space $Y$, loops can enclose the hole
any number $n \in \Z$ times and one may regard clockwise orientation as positive $n$
and anticlockwise orientation as negative $n$. There is an infinite number of homotopy
classes of loops in $Y$, each associated with this winding number $n$.

Loops can be combined (multiplied or added). For example $\alpha * \beta $ is taken to mean traversal 
firstly of loop $\alpha$ and subsequently of loop $\beta$. Inverse loops can also be defined, 
so that $\alpha^{-1}$ has the same location but reverse orientation to $\alpha$. The product 
$\alpha * \alpha^{-1}$ is then homotopic to (not equal to) 
the identity loop. 

In fact, since a loop $\alpha$ is homotopic to an infinite number of other loops, 
it is more convenient to consider just one loop, representative of that homotopy class, or
better, to consider the homotopy class itself -- which we label $[\alpha]$. 
Products of classes are defined as classes of products:
$
 [\alpha] * [\beta ] = [\alpha * \beta ]
$,
and, in particular, $[\alpha] * [\alpha^{-1}] $ is equal to (not merely homotopic to) 
the homotopy class of the identity loop.
The set of homotopy classes defined in this manner has the properties of
closure, asociativity, the existence of an inverse and an identity. Therefore it forms a group. 
\begin{defn}
\label{defn14}
The group of homotopy classes of loops in a topological space $X$ based at a point $x_0$ 
is denoted by $\pi_1(X,x_0)$ and is called the fundamental group or first homotopy group. 
If $[\alpha ], [\beta] \in \pi_1(X,x_0)$, then their product is defined as  
$[\alpha] * [\beta ] = [\alpha * \beta ]$. The identity element is the class
of all loops homotopic to the degenerate loop comprised soley of the point $x_0$.
\end{defn}

At this point, we remark that, in the general definition~\ref{defn14}, the fundamental group
is seen to depend on the base point $x_0$. This rather cumbersome burden disappears 
if one restricts one's considerations to pathwise-connected topological spaces. 
A space $X$ is pathwise-connected  (also called path-connected or 0-connected) 
if every pair of points $x_0,x_1 \in X$ are connected by a path 
$\gamma$ (i.e., $\gamma:[0,1] \rightarrow X$
such that $\gamma(0)=x_0$ and $\gamma(1) = x_1$).
This is actually a stronger concept than that of connectedness in definition~\ref{defn11}, i.e.,
connected spaces exist which are not pathwise-connected. For the topological 
spaces encountered herein the two concepts coincide.

Before stating the theorem establishing the redundancy of a base point 
in considerations of the fundamental group structure of most useful topological spaces, we recall 
some basic concepts from group theory.
\begin{defn}
An Abelian group is one in which the elements commute, i.e., if $G$ is an Abelian group,
for $a, b \in G$, $a b = ba$. 
\label{defn15}
\end{defn}
For example, $\Z_n$ is the cyclic group generated by a single element which satisfies $g^n=0$
where $0$ is the identity element. 
All cyclic groups are Abelian.
\begin{defn}
A group homomorphism $f$ between two groups $G$ and $H$ is a map which preserves the group structure, i.e.,
for $a, b \in G$, $f(a b) = f(a) f(b)$. 
\label{defn16}
\end{defn}
This definition is sufficient to ensure that
the identity is mapped to the identity and that the inverse map is preserved.
A counter-example in the group of reals under multiplication is  the sine function,
which is not a homomorphism because $\sin{(a b)} \ne \sin{a} \sin{b}$.
\begin{defn}
\label{defn17}
 A bijective map $f$ is both injective (also called one-to-one and for which
 $a\ne b \Rightarrow f(a) \ne f(b)$) and surjective (also called onto and for which
 $\forall b \in H$, $\exists$ $a \in G$ s.t. $f(a) = b$). 
\end{defn}
\begin{defn}
\label{defn18}
 A bijective homomorphism between groups $G$ and $H$ is called an isomorphism. We write
 $G \cong H$.
\end{defn}
So a homeomorphism is a continuous isomorphism. (The word `isomorphism' 
is derived from the Greek,
meaning equal shape.)
\begin{thm}
 If $x_0,x_1$ belong to the pathwise-connected topological space $X$, then 
 $\pi_1(X,x_0) \cong \pi_1(X,x_1)$.
\label{thm1}
\end{thm}

This theorem establishes the independence of $\pi_1(X,x_0)$ from $x_0$ for reasonable 
topological spaces. 
We now seek to address the question to what extent does the 
fundamental group depend on $X$ itself. It turns out that it is also independent of $X$
up to homeomorphism. I.e., if $X$ is homeomorphic to  $Y$, then $\pi_1(X,x_0)$ is essentially 
the same as (isomorphic to) $\pi_1(Y,y_0)$. In fact one can do better than that, but we
need yet another new concept, namely homotopy type, which is broader than homeomorphism.
It is given by relaxing the equality in definition~\ref{defn12} to homotopy. 
\begin{wrapfigure}{i}{0.5\textwidth}
\vspace{-2.0cm}
\hspace{2.8cm}
\centerline{\includegraphics[width=0.75\textwidth]{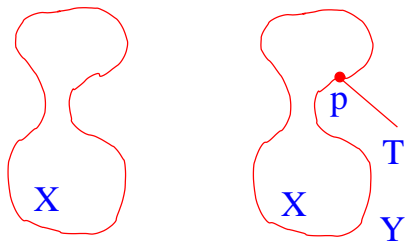}}
\vspace{-11.0cm}%
\caption{Spaces $X$ and $Y$ (which consists of a tail $T$ glued on to $X$ at the point $P$)
are of the same homotopy type but are not homeomorphic. }
\vspace{-0.5cm}
\label{Fig:homotopytype}
\end{wrapfigure}

\begin{defn}
\label{defn19}
 Topological spaces $X$ and $Y$ are said to be 
homotopy equivalent or of the same homotopy type 
if there exist continuous maps 
$f: X \rightarrow Y$ and $g: Y\rightarrow X$ such that 
\[
 f \circ g \simeq 1_Y\;, \quad \quad g \circ f \simeq 1_X\;.
\]
\end{defn}

So two homeomorphic spaces are necessarily of the same homotopy type, but the converse is
not  the case. Refer to figure~\ref{Fig:homotopytype} where
the space $Y$ is formed from space $X$ by connecting a `tail', as illustrated, and let 
$f(x) = x$, and $g(y)=y$ if $y \in X$, while $g(y) = P$ if $y$ is in the tail. 
Clearly $X$ and $Y$ are not homeomorphic (since $f \circ g \ne 1_Y$ in the tail), 
but they are of the same homotopy type.
Thus sets of the same homotopy type have the same essential structure --
they are the same up to stretching, twisting or compression but not under cutting.
\begin{thm}
 If the pathwise-connected topological spaces $X$ and $Y$ are of the same homotopy type, then 
\[ 
 \pi_1(X,x_0) \cong \pi_1(Y,y_0) \;.
\]
\label{thm111}
\end{thm}
In particular, if $X$ and $Y$ are homeomorphic, then their fundamental groups 
are isomorphic: $\pi_1(X,x_0) \cong \pi_1(Y,y_0) $.
Note that the converse does not necessarily hold.
Nonetheless, this establishes the central result that the 
fundamental group is a topological invariant of a space.

The fundamental group plays a critical role in the classification of topological defects 
in statistical physics and beyond. Fortunately it is possible to determine this group 
in most cases. To do that, we need the following definition.
\begin{defn}
\label{defn20}
Let $R \subset X$. If there exists a continuous map $r: X \times [0,1] \rightarrow X$, such that
\[
r(x,0) = x \quad \forall \quad x \in X\;, 
\quad \quad
r(x,t) = x \quad \forall \quad x \in R\;,
\quad \quad 
r(x,1) \in R\;,
\]
then $R$ is called a deformation retract of $X $ and the map $r$ is called a retract.
\end{defn}

\begin{wrapfigure}{i}{0.5\textwidth}
\vspace{-3.5cm}
\hspace{2.8cm}
\centerline{\includegraphics[width=0.75\textwidth]{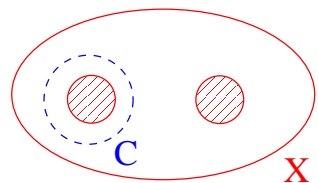}}
\vspace{-10.5cm}%
\caption{The circle $C$ (dashed) is not a deformation retract of the 
full space $X$.}
\label{Fig:defretract}
\vspace{-0.6cm}%
\end{wrapfigure}

The deformation retract is a subspace of an original space, formed by
continuous shrinking. In figure~\ref{Fig:homotopytype}, the space $X$ is a deformation
retract of space $Y$. Similarly an annulus, for example can be retracted into a circle.
A counter example illustrated in figure~\ref{Fig:defretract}, 
where the circle $C$ is not a deformation retract of the full space $X$
because the hole on the right is an obstruction to such a retraction process.
The usefullness of this lies in the following theorem:

\begin{thm}
 If $R$ is a deformation retract of a pathwise connected topological space $X$, then 
$\pi_1(X,x_0) \cong \pi_1(R,x_0)$.
\label{thm3}
\end{thm}
Deformation retracts can be used to construct a
representative of a space, called a {\emph{polyhedron}}, 
which contains all the homotopic features of that space. One can then use an algorithm to 
calculate its fundamental group. This algorithm is illustrated in the following examples.\\

\begin{wrapfigure}{i}{0.5\textwidth}
\vspace{-3.5cm}
\hspace{2.0cm}
\centerline{\includegraphics[width=0.7\textwidth]{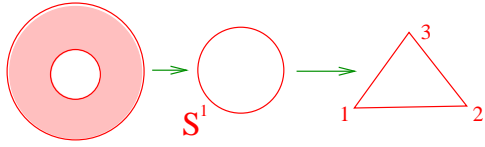}}
\vspace{-10.cm}%
\caption{Triangulation of an annulus.}
\label{Fig:annulus}
\end{wrapfigure}

\noindent
{\bf{Example~1 (the annulus)}}
To determine the fundamental group of an annulus, one firstly
deforms it to a circle $S^1$ and then  {\emph{triangulates}} that
space using {\emph{simplices}} (see figure~\ref{Fig:annulus}).
A $0$-simplex is a point, a $1$-simplex is a line segment, 
a $2$-simplex is a triangle (including its interior), a $3$-simplex is a tetrahedron
(including its interior), and so on. Here we have three $0$-simplices 
($\{1\}$, $\{2\}$ and $\{3\}$) and three $1$-simplices 
($\{1,2\}$, $\{2,3\}$ and $\{3,1\}$). The $2$-simplex $\{1,2,3\}$ is not
involved as the cavity within the triangle is not part of the space 
(the annulus, by definition, has a hole).
Each $1$-simplex corresponds to a group element, which 
we label $g_{12}$, $g_{23}$ and $g_{31}$.
Next, construct a `scaffold' or {\emph{contractable subpolyhedron}} which 
contains all the vertices of the triangulation or polyhedron.
In our case $\{1,2\} \cup \{2,3\}$ will suffice. 

\begin{wrapfigure}{i}{0.5\textwidth}
\vspace{-5.cm}
\hspace{4.8cm}
\centerline{\includegraphics[width=0.9\textwidth]{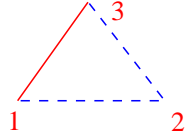}}
\vspace{-12.5cm}%
\caption{A subpolyhedron (dashed) spanning the triangulation of $S^1$.}
\label{Fig:tri}
\end{wrapfigure}

This contractable subpolyhedron is illustrated for the present example in
figure~\ref{Fig:tri}. Each of the $1$-simplices in the spanning
subpolyhedron gives the identity group element which we denote by $0$, 
so $g_{12} = g_{23} = 0$.
We are left with one non-trivial element, namely $g_{31} \equiv g$.
The group generated by such an element is $\Z$, the integers under addition.
Therefore the fundamental group of the annulus is $\pi_1(S^{1})=\Z$.
\\

\noindent
{\bf{Example~2 (the disc)}} The disc $D$ or $D^2$ is again triangulated by a triangle,
but this time its interior is included. This is the $2$-simplex $\{1,2,3\}$.
Such a $2$-simplex gives a relation $g_{12}g_{23}g_{31}^{-1} = 0$, which,
with $g_{12} = g_{23} = 0$ gives $g_{31} = 0$. So all elements of the group are 
the identity and $\pi_1(D) \cong \{0\}$.\\

\begin{wrapfigure}{i}{0.5\textwidth}
\vspace{-4.0cm}
\hspace{2.8cm}
\centerline{\includegraphics[width=0.8\textwidth]{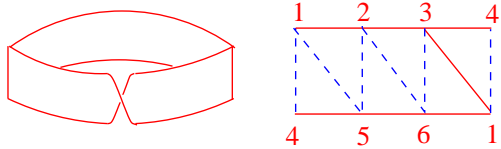}}
\vspace{-11.3cm}%
\caption{The M{\"o}bius strip (left), its triangulation (right) and a spanning subpolyhedron (dashed).}
\label{Fig:Moebius}
\end{wrapfigure}

\noindent
{\bf{Example~3 (the M{\"o}bius strip)}} 
The M{\"o}bius strip, a suitable triangulation and a contractable subpolyhedron
are illustrated in figure~\ref{Fig:Moebius}. Each of the five $1$-simplices
contained in the subpolyhedron give the identity group element. There are seven
remaining group elements corresponding to the seven $1$-simplices outside
the subpolyhedron (taking care to note the left and right edges are the same).
But there are six $2$-simplices giving six relations between these seven elements.
Only one non-trivial element remains and $\pi_1({\mbox{M{\"o}bius}}) \cong \Z$. 

For any of these simple spaces, we can guess the fundamental group by imagining 
whether or not loops straddling the structure can be deformed to a point.

\begin{wrapfigure}{i}{0.5\textwidth}
\vspace{-3.6cm}
\hspace{4.2cm}
\centerline{\includegraphics[width=0.8\textwidth]{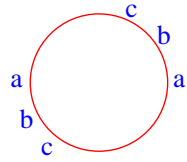}}
\vspace{-11.3cm}%
\caption{Formation of the projective plane RP$^2$ by identification of antipodal points 
on a disc.}
\label{Fig:RP2}
\end{wrapfigure}

~\\
\noindent
{\bf{Example~4 (the projective plane)}} The real projective plane RP$^2$ is constructed
by identifying each pair of diametrically opposed points on the boundary of a disc
(figure~\ref{Fig:RP2}).
A path connecting a point to its antipode is therefore closed and constitutes a loop
which cannot be shrunk to a point. A trivial loop is only constructed by 
returning to the starting point (not antipode). So, besides the identity, 
there is a nontrivial group element, which is its own inverse and
$\pi_1({\rm{RP}}^2)  \cong \Z_2 \equiv \{0,g\}$.

To determine the fundamental group of higher-dimensional spaces,
van Kampen's  theorem is useful:
\begin{thm}
\label{thm4}
 If $X$ and $Y$ are topological spaces, the fundamental group of their
product is the direct product of their fundamental groups.
\[
 \pi_1(X \times Y) \cong \pi_1(X) \oplus \pi_1(Y) \;.
\]
\end{thm}
For example, the three-dimensional torus is given by $T^2=S^1 \times S^1$ so 
$\pi_1(T^2) = \pi_1(S^1) \oplus \pi_1(S^1) = \Z \oplus \Z \equiv \Z^2$.

\section{The higher homotopy groups}
\label{higherhomotopy}

Only $1$- and $2$-simplices are used in the algorithm to calculate the fundamental
(or first homotopy) group.
This means that $\pi_1$ can only detect two-dimensional holes present in a space
(or combinations thereof, such as the two two-dimensional holes in the torus).
For example, by considering shrinkage of a loop on its
surface, it is easy to convince oneself that the fundamental group 
of a sphere in $3$-space is trivial: $\pi_1(S^2) \cong \{0\}$.
Such one-dimensional loops are incapable of detecting the three-dimensional interior of the 
sphere. To deal with this circumstance we need higher homotopy groups.

While the reader is referred to the literature for the mathematial details \cite{NaSe88}, 
the basic idea is straightforward and now outlined. 
The $1$-loops we encountered hitherto can be thought of as elastic bands tied at a 
basepoint $x_0$.
The two-dimensional equivalent of such an object can be thought of as a balloon, with the neck
anchored at $x_0$. Such a $2$-loop can encompass a spherical hole in much the same was as
a $1$-loop can enclose a circular one. Moreover, a balloon can be wrapped or unwrapped
an integer number
of times around a sphere and a group structure can be built up in an equivalent manner
to the (homotopy equivalent classes of) $1$-loops.
In still higher dimensions, $n$-loops can be defined. 

\begin{table}
\caption{The homotopy groups of the $n$-sphere. The higher groups are Abelian but fall into
no easily distinguishable pattern.}  
\begin {center}
\begin{tabular}{cr|ccccccccc} 
  \\
\hline \hline
$\pi_m(S^n)$ &   &   &  &   &  &  &  &  & &  \\
 &   & \multicolumn{8}{l}{$n$}  \\
 $m$ &&    1    &    2     &   3      &    4     &    5    &    6   &     7      &      8      & $\dots$ \\
\hline 
 &    &         &           &          &         &         &         &            &            &  \\
 & 0  & $\{0\}$ & $\{0\}$  & $\{0\}$  & $\{0\}$ & $\{0\}$ & $\{0\}$ & $\{0\}$   & $\{0\}$   &$\dots$ \\
 & 1  & $\Z$    & $\{0\}$  & $\{0\}$  & $\{0\}$ & $\{0\}$ & $\{0\}$ & $\{0\}$   & $\{0\}$   &$\dots$ \\
 & 2  & $\{0\}$ & $\Z$     & $\{0\}$  & $\{0\}$ & $\{0\}$ & $\{0\}$ & $\{0\}$   & $\{0\}$   &$\dots$ \\
 & 3  & $\{0\}$ & $\Z$     & $\Z$     & $\{0\}$ & $\{0\}$ & $\{0\}$ & $\{0\}$   & $\{0\}$   &$\dots$ \\
 & 4  & $\{0\}$ & $\Z_2$   & $\Z_2$   & $\Z$    & $\{0\}$ & $\{0\}$ & $\{0\}$   & $\{0\}$   &$\dots$ \\
 & 5  & $\{0\}$ & $\Z_2$   & $\Z_2$   & $\Z_2$  & $\Z$   & $\{0\}$  & $\{0\}$   & $\{0\}$   &$\dots$ \\
 & 6  & $\{0\}$ & $\Z_{12}$& $\Z_{12}$& $\Z_2$  & $\Z_2$ & $\Z$    & $\{0\}$   & $\{0\}$   &$\dots$ \\
 & 7  & $\{0\}$ & $\Z_{2}$ & $\Z_{2}$ &$\Z\oplus\Z_{12}$&$\Z_2$&$\Z_2$&$\Z$   & $\{0\}$   & $\dots$ \\
 & 8  & $\{0\}$ & $\Z_{2}$ & $\Z_{2}$ &$\Z_2\oplus\Z_{2}$&$\Z_{24}$&$\Z_2$&$\Z_2$&$\Z$ &$\dots$ \\
 & 9  & $\{0\}$ & $\Z_{3}$ & $\Z_{3}$ &$\Z_2\oplus\Z_{2}$&$\Z_{2}$&$\Z_{24}$&$\Z_2$&$\Z_2$ &$\dots$ \\
 &    & $\vdots$&$\vdots$ & $\vdots$ &$\vdots$ & $\vdots$& $\vdots$&$\vdots$& $\vdots$& \\
 &    &         &           &          &          &         &           &         &  & \\
\hline \hline
\end{tabular}
\end{center}
\label{tab1}
\end{table}

\begin{defn}
\label{defn21}
 The homotopy class of $n$-loops in a topological space $X$ with basepoint $x_0$ 
forms the $n$-dimensional homotopy group $\pi_n(X,x_0)$.
\end{defn}
It turns out that, while some but not all fundamental groups are Abelian, {\emph{all}}
higher homotopy groups are Abelian.
While there is no algorithm to calculate the higher homotopy groups comparable to
the case of the fundamental group, 
nor is there a higher homotopy analogue to van Kampen's theorem,
there are tools such as 
Freudenthal's theorem: 
\begin{thm}
\label{thm5}
  The $m$-dimensional homotopy group of the $n$-sphere 
 $\pi_m(S^n)$ depends only on $m-n$ for $m \le 2(n-1)$. 
\end{thm}
An immediate consequences of this is
$\pi_m(S^n) \cong \pi_{m+1}(S^{n+1})$ for $m \le 2(n-1)$. Further, since 
$\pi_1(S^n) \cong \{0\}$ for $n \ge 2$,
$\pi_m(S^n) \cong \{0\}$ for $n \ge m+1$.
If $n=m$, one has (for $n \ge 2$) $\pi_n(S^n) \cong \pi_2(S^2) \cong \Z$. 

Finally,  although $\pi_0$ is not a group, it is often used to denote the number of 
connected domains or components in a space. 

Classification of the homotopy groups of topological spaces is an active field
of research and has generated many surprising results.
Table~\ref{tab1} illustrates this by listing $\pi_m(S^n)$ for small $m$ and $n$. 
While there is no overall identifiable pattern, as the dimension increases 
some regularity does occur. 
In particular, $\pi_{n+1}(S^n)=\Z_2$ for $n \ge 3$,
 $\pi_{n+2}(S^n)=\Z_2$ for $n \ge 2$,
 $\pi_{n+3}(S^n)=\Z_{24}$ for $n \ge 5$ and so on.

Homotopy groups for other topological spaces have also been determined and some of those
more commonly used in physics are listed 
in table~\ref{tab2}.
\begin{table}
\caption{The homotopy groups of various useful topological spaces.}  
\begin {center}
\begin{tabular}{l} \hline \hline
    \\
{$ \pi_1(\R^n) \cong \pi_1(D) \cong \pi_m(S^{n>m}) \cong \pi_{m>2}(S^1) \cong \{0\}$}  \\
   $\pi_1(T^2) \cong \Z \oplus \Z \equiv \Z^2$     \\
   $\pi_1(T^n) \cong \Z^n$    \\
  $\pi_1({\rm{RP}}^n) \cong  \Z_2$ ($n\ge 2$)  \\
  $\pi_m({\rm{RP}}^n) \cong \pi_m(S^n)$ for $m \ge 2$   \\
  $\pi_n({\rm{RP}}^n) \cong \Z$     \\
\hline \hline
\end{tabular}
\end{center}
\label{tab2}
\end{table}

This completes the pedagogic introduction to homotopy. Equipped with the above topological notions,
one may now examine the role played by homotopy  in statistical physics.

\section{Phase transitions and topological defects}
\label{phasetransitions}

Phase transitions are amongst the most ubiquitous and remarkable phenomena in nature and involve
abrupt or gradual changes in quantifiable macroscopic properties 
brought on by varying a system's parameters such as temperature, pressure or a coupling.
Equilibrium statistical physics which describes such phenomena is based on the premise
that the probability 
that a system is in a state $S$ with energy $E$ at a temperature $T$ is
\begin{equation}
 P(S) = \frac{\exp{(- \beta E (S) )}}{Z_L(\beta)} \,,
\label{one}
\end{equation}
where  $\beta = 1/ k_B T$, $k_B$ is a universal constant, and $Z_L(\beta)$ is 
the partition function which serves as a normalising factor. It is given by
\begin{equation}
 Z_L(\beta) = \sum_S \exp{(- \beta E(S))}
\,.
\end{equation}
Here, and in the following, the linear extent of the system is indicated by the subscript $L$.
Another fundamental quantity is the free energy, $f_L(\beta)$, given by
\begin{equation}
 f_L(\beta) = \frac{1}{L^d}\ln{Z_L(\beta)}
\;,
\end{equation}
where $d$ is the dimensionality of the system so that its volume is $L^d$.
In the modern classification scheme, phase transitions are categorised as first, second or higher
order if the lowest derivative of the infinite-volume free energy $f_\infty(\beta)$ 
that displays non-analytic behaviour is the 
first, second or higher one (see, e.g., \cite{JaJo06}). 

For an $O(n)$-symmetric model, the sites $i$ of a $d$-dimensional lattice are 
occupied by $n$-dimensional unit-length vectors $\vec{s}_i \in S^{n-1}$ which 
may be considered to
reside in an internal spin space. 
In general then, $n$ may be independent of the dimensionality $d$ of the 
physical space or medium.
For the $O(n)$ model, the energy of a given configuration
is given as
\begin{equation}
 E = - \sum_{\langle i, j \rangle }{\vec{s}_i\vec{s}_j}
\;, 
\label{XY}
\end{equation}
where the summation runs over nearest neighbouring sites or links of the lattice.

The $n=1$ version of this construct is the Ising model.
If $n=2$, the spins live in a plane and the model is referred to as the $XY$ model.
The $O(3)$ version is called the Heisenberg model and the limit $n \rightarrow \infty$
corresponds to the spherical model.
In each of these models, the system (and the partition function in particular)
is invariant under rotations in spin space. 
However, the order parameter or spin expectation value $\langle \vec{s}_i \rangle $
may, in principle, not respect this symmetry. This very common circumstance is referred to as 
spontaneous symmetry breaking.
Let $T_c$ or $\beta_c$ denote the phase transition point.
The conventional scaling scenario at such a transition is of the power-law type in 
 the reduced temperature $t=T/T_c-1$. For example, this is the type of scaling which 
characterises the standard temperature-driven paramagnetic-ferromagnetic
phase transition in the  $XY$ model in $d>2$ dimensions.

\begin{wrapfigure}{i}{0.5\textwidth}
\vspace{-2.0cm}
\hspace{3.2cm}
\centerline{\includegraphics[width=0.8\textwidth]{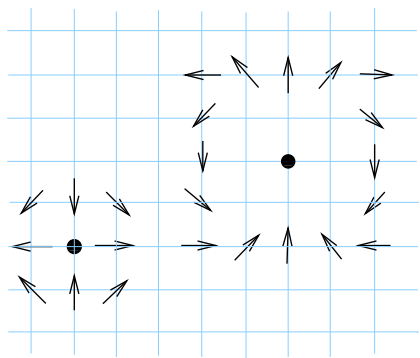}}
\vspace{-11.0cm}%
\caption{Examples of an antivortex (or vortex of charge $-1$) 
and a vortex (of charge $2$) in the $XY$ model on a regular lattice. 
As contours around the two bold points are traversed (counterclockwise), 
spins rotate through $-2\pi$ and $4\pi$ respectively.}
\label{Fig:vortices}
\end{wrapfigure}

However, the existence of a phase with long-range order is precluded in two-dimensional models
with continuous symmetry group and continuous interaction Hamiltonian, such as (\ref{XY}), 
due to the Mermin-Wagner theorem \cite{MW}.
Nonetheless, a temperature-driven transition can still exist -- to a phase with topological order.
The two-dimensional $XY$ model has a famous infinite-order phase transition 
of this type which breaks no system symmetries and is 
called the Berezinskii-Kosterlitz-Thouless (BKT) transition \cite{B,KT}. 
The transition is mediated by topological defects called vortices. 
Such vortices can be identified by tracking the well-defined values of the spins $\vec{s}_i$
along a given contour  (see figure~\ref{Fig:vortices}). 
Spins can rotate along such
a contour through $2n\pi$ for
$n\in \Z$. If $n \ne 0$, one speaks of a vortex (actually one may refer
to a negative-$n$ vortex as an antivortex). 
In the absence of the lattice, shrinking the contour to a point would lead to a singularity
in the presence of such a vortex. 

From homotopy theory, it is clear that the winding number in the above two-dimensional 
example is 
related to the fundamental group of the order-parameter space $\pi_1(S^1)$.
In principle, in a $d=2$ dimensional medium, one could also have a line 
defect 
if $\pi_0$ was non-trivial (since $\pi_0$ denotes the number of connected domains
in the medium, such a defect is called a domain wall).
In a $d=3$ 
medium, defects 
of dimension $0$, $1$ or $2$ may exist.
These are point defects (monopoles), 
           line defects (vortices)
or         wall defects (domain walls), 
and are detected by $\pi_2$, $\pi_1$ and $\pi_0$ respectively.
In  general, then, an $m$-dimensional defect in a $d$ dimensional 
medium is classified by $\pi_{d-m-1}(S)$, where $S$ is the order-parameter space. 
They are called domain walls (classified by $\pi_0$),
vortices ($\pi_1$), monopoles ($\pi_2$), instantons ($\pi_3$), etc. 
The classification scheme is summarized in table~3.

\begin{table}
\begin{center}
\caption{Classification of topological defects. Here $d$ is the dimensionality of the medium and 
$m$ is that of the defect. }  
\vspace{0.7cm}
\begin{tabular}{l|lllll}
\hline\hline\\
        & $m=0$ & $m=1$ & $m=2$ & $m=3 \dots$ \\ 
\hline
 $d=1$  & { $\pi_0$} &                          &                         & &  \\ 
 $d=2$  & {   $\pi_1$} & { $\pi_0$}  &                         & &  \\ 
 $d=3$  & {  $\pi_2$} & {   $\pi_1$}  & { $\pi_0$} & &   \\ 
 $d=4$  & {  $\pi_3$} & {   $\pi_2$} & {   $\pi_1$} & { $\pi_0$} &      \\ 
$\quad \vdots$    &   & 
$\!\!\!\!$ 
{\rotatebox{-18}{instantons} } 
& 
$\!\!\!\!\!\!\!\!\!\!$
{\rotatebox{-18}{monopoles}}
&
$\!\!\!\!\!\!\!$
{ \rotatebox{-18}{vortices}}
& 
$\!\!\!\!\!\!\!\!\!\!$
{ \rotatebox{-18}{domain walls} }     \\ 

\end{tabular} 
\end{center}
\label{tab33}
\end{table}

\section{The  
BKT transition in the $XY$ and step models}
\label{KT}

The $XY$ model in two dimensions 
is the generic one for the study of the topo\-logical\-ly-mediated BKT transition.
It is also used to study systems such as films of superfluid helium, Josephson-junctions, 
superconducting materials, fluctuating surfaces as well as certain magnetic, 
gaseous and liquid-crystal systems.
Besides this model, transitions of the BKT type also exist in 
the ice-type $F$ model \cite{WeJa05},
antiferromagnetic models \cite{CaVa98}, 
certain models with long-range interactions \cite{Ro97}, 
lattice gauge theories \cite{WeBi05}
and in string theory \cite{Ma02} amongst others. 
Thus a thorough and quantitative understanding 
of the paradigmatic two-dimensional $XY$ model is beneficial to a 
number of areas within theoretical physics.
The BKT transition in this model has come under intense scrutiny, and only in recent years has 
the picture begun to become clear. 

At low temperature, vortices and antivortices appear in pairs
and there is quasi-long-range order (meaning orientation of spins over small length scales).
The correlation function characterises this ordering and, with $x_i$ denoting the position
of the $i^{\rm{th}}$ site, its leading critical behaviour from
renormalization group (RG) theory is \cite{KT}
\begin{equation}
\label{correlationfunction}
G_\infty (x_i-x_j)= \langle s_i s_j \rangle \sim |x_i-x_j|^{-(d-2+\eta_c)}
\end{equation}
(with $d=2$ in this instance).
The correlation length in an infinitely large system is denoted by $\xi_\infty(t)$ and
measures to what extent 
spins at different sites are correlated. It diverges in the  
massless low-temperature phase. This divergence persists, with the system remaining critical with
varying $\eta(\beta)$, up to $\beta = \beta_c$ at which $\eta(\beta_c) = \eta_c = 1/4$.

According to the BKT picture, the number of vortex pairs increases as the temperature 
is raised and they proliferate in the high-temperature phase.
At the transition point, it becomes no longer sensible to describe vortices as belonging to paired sets
and they are said to unbind. In the high-temperature phase, the dissociation of vortices destroys
the order of the system.
Above this  point, correlations decay exponentially fast 
with leading behaviour
\begin{equation}
 G_\infty(x_i-x_j) \sim \exp{\left[-|x_i-x_j|/\xi_\infty(t)\right]}
\,.
\end{equation}
The leading 
approach to the transition from the high-temperature phase is characterised by essential singularities,
\begin{eqnarray}
 \xi_\infty (t) & \sim & \exp{\left(bt^{-\nu}\right)} \; , 
\label{a}
\\
 C_\infty  (t)  & \sim & \xi_\infty^{-2} \; , 
\label{b}
\\
 \chi_\infty (t) & \sim & \xi_\infty^{2-\eta_c} \; ,
\label{c}
\end{eqnarray}
for the correlation length, the specific heat and the susceptibility
(which respectively measure the response of the system to variations in the temperature 
and application of an external magnetic field). Here $b$ is a non-universal constant.

The `BKT scenario' is thus  taken to mean a phase transition which 
(i)  exhibits essential scaling behaviour 
and
(ii) is mediated by a vortex-binding mechanism.
This picture is based on perturbation theory \cite{B,KT} 
and a variety of techniques have been employed
for over thirty years in attempts to verify it.

\subsection{BKT scaling behaviour}

These attempts at verification have included 
non-perturbative simulational and high-temperature series analyses of the $XY$ model.
In particular, verification of the analytical BKT RG
prediction that $\nu = 1/2$
and $\eta_c = 1/4$ and accurate determination of the value of $\beta_c$ proved elusive. 
Typically the critical temperature was determined by firstly fixing 
$\nu=1/2$. However, measurements of $\eta_c$ then yielded a value
incompatible with the BKT prediction. 
The elusiveness of such unambiguous corroborative evidence led to the 
essential nature of the transition in the $d=2$ $XY$ model being  questioned 
\cite{SeSt88Kim}. 
An extensive overview of the status of the model up to 1997
is contained in  \cite{KeIr97}.

A self-consistency analysis, based on Lee-Yang zeros, was used in \cite{KeIr97,KeIr96} 
to show that the thermal scaling forms
(\ref{b}) and (\ref{c}) (and similar formulae for other quantities) 
are mutually incongruent as they stand, and have to 
be modified to include multiplicative logarithmic corrections. In fact,
\begin{eqnarray}
 C_\infty (t) &  \sim & \xi_\infty(t)^{-2} \left( \ln{\xi_\infty(t)} \right)^{\tilde{q}}\,,
\label{ca}
\\
\chi_\infty (t) &  \sim & \xi_\infty(t)^{2-\eta_c} \left( \ln{\xi_\infty(t)} \right)^{-2r} \,,
\label{cc}
\end{eqnarray}
and
\begin{equation}
G_\infty(x_i-x_j)   \sim |x_i-x_j|^{-\eta_c}\left( \ln{|x_i-x_j|}\right)^{-2r}\,,
\label{Ga}
\end{equation}
with  ${\tilde{q}}=6$ and 
where  RG  indications implicit in \cite{KT}
are that $r=-1/16 = -0.0625 $. Besides this implication, 
the first indication of the existence of a non-zero correction exponent $r$ 
was presented by Butera and Comi  using high-temperature series expansions \cite{BuCo93}.
On the other hand, arguments that BKT theory implies 
the absence of multiplicative logarithmic corrections, i.e. $r=0$,
have also been recently made~\cite{Ba01,Pa01}.

\noindent
\begin{table}[th]
\caption{Estimates for the critical point $\beta_c$ for the $XY$ model 
as well as the logarithmic-correction exponent $r$ for the BKT universality class
from a selection of recent papers.
}
\begin{center}
\begin{small}
\vspace{0.7cm}  
\noindent\begin{tabular}{|l|l|l|l|l|} 
    \hline     \hline
      & & & &  \\
     Authors  &$\!\!\!$ Year  & $\!\!\!$ Method $\!\!\!\!$ & $\beta_c$ & $r$  \\
      & & & &  \\
    \hline
Kosterlitz \& Thouless    \cite{KT}     &1973 & RG     &            &  $-0.0625$        \\
Irving \& Kenna           \cite{KeIr97,KeIr96} &1996 & FSS    & $1.113(6)$ &  $-0.02(1)$       \\ 
Patrascioiu \& Seiler     \cite{PaSe96} &1996 & thermal&            & $\quad 0.077(46)$\\
Campostrini, Pelissetto,  &1996 & thermal& $1.1158(6)$&  $\quad0.042(5)$ \\ 
Rossi \& Vicari   \cite{CaPe96} &     &        & $1.120(4)$ &  $\quad0.05(2)$  \\
Janke                     \cite{Ja97} &1997 & FSS,    &            & $ -0.027(1)$       \\
                                       &     & thermal&            &  $\quad0.0560(17)$ \\
Hasenbusch \&  Pinn   \cite{HaPi97}    &1997 & RG     & $1.1199(1)$&                     \\
Jaster \&  Hahn    \cite{JaHa99} &1999 & FSS,    &            & $-0.0233(10)$\\
                                 &     & thermal &            & $\quad0.056(9)$/$0.070(5)$  \\
Balog, Niedermaier, Niedermayer,   &2001 & RG,    &            & $\quad 0$\\
Patrascioiu, Seiler, Weisz   \cite{Ba01,Pa01}    & &  thermal    &            & \\
Tomita \& Okabe  \cite{ToOk02}       &2002 & FSS    & $1.1194(8)$ & $\quad 0.038(5)$                        \\
Dukovski, Machta \& Chayes  \cite{DuMa02}       &2002 & FSS    & $1.120(1)$ &                         \\
Chandrasekharan \&   Strouthas \cite{ChSt03}          &2003 & FSS    &             & $-0.035(10)$       \\
Hasenbusch                \cite{Ha05} &2005 & FSS    &             & $-0.056(7)$        \\
 & & & & \\
    \hline
    \hline 
  \end{tabular}
\end{small}
\end{center}
\label{tab4}
\end{table}
The formulae (\ref{ca}) and (\ref{cc}) refer to thermal scaling on infinite lattices. Such systems
are not amenable to computational techniques as one may only simulate
finite-size systems. There, finite-size scaling (FSS) theory predicts that the
role of the correlation length is played by the lattice extent $L$. 
For example, the susceptibility at criticality ($t=0$) scales as \cite{KeIr96}
\begin{equation}
\chi_L (0)  \sim L^{2-\eta_c} \left( \ln{L} \right)^{-2r} \; .
\label{ccFSS}
\end{equation}
Verification of the BKT scaling scenario then requires numerical confirmation of the thermal 
formula (\ref{cc}) and/or the FSS formula (\ref{ccFSS}).
In \cite{KeIr97,KeIr96}, the hitherto conflicting
results for $\nu$,  $\beta_c$ and $\eta_c$ were finally resolved. 
However, the analysis resulted in an estimate of $-0.02(1)$
for $r$, a value in conflict with the RG 
prediction of $r=-1/16 = -0.0625 $ from \cite{KT}.

This prompted the shifting of the focus of recent numerical 
studies of the $XY$ model to the determination of the logarithm exponent $r$.
Indeed, in \cite{Ja97,JaHa99,ChSt03}, FSS analyses using (\ref{ccFSS}),
yielded values compatible with 
that of \cite{KeIr97,KeIr96} but incompatible with \cite{KT} (see table~4 
and \cite{BeSh04} for an overview). 
Nonetheless, it was clear that  the resolution of the $\nu$-$\beta_c$-$\eta_c$ controversy
is achieved by taking the logarithmic corrections into account. 
The most precise numerical estimate for the critical temperature 
for the $XY$ model is $\beta_c = 1.1199(1)$ \cite{HaPi97}, a value obtained 
by mapping the model onto the  body-centered solid-on-solid model (which is exactly solvable), 
and thereby circumventing the issue of logarithmic corrections. 
Indeed, the recent analyses which have included logarithmic 
corrections have resulted in estimates for
$\beta_c$ compatible with this value and with $\eta = 1/4$ for the anomalous dimension.

The study of the $XY$ model contained in \cite{Ja97} employs the Villain formulation, which
has an action different to (\ref{XY}) and leads to a different value for quantity $\beta_c$
(which is nonuniversal). Similarly, in table~4, 
\cite{JaHa99} and \cite{ChSt03} contain studies
of a lattice grain boundary model and a lattice gauge theory respectively.
Although the critical temperatures for these models bear no relationship to their $XY$ counterpart,
they have the same scaling behaviour as they are in the same universality class.
Thus the value for the logarithmic exponent $r$ should be the same as in the $XY$ model. 

It was further suggested in \cite{KeIr97,Ja97} that since taking the logarithmic corrections into account 
leads to the resolution of the leading-scaling puzzle,
in the same spirit 
numerical measurements for $r$ may become compatible with the RG prediction if
sub-leading scaling corrections are taken into account. 
In this case, (\ref{ccFSS}) is more fully expressed as
\begin{equation}
\chi_L (0)  \sim L^{2-\eta_c} \left( \ln{L} \right)^{-2r} 
\left\{
 1 + {\cal{O}} \left( \frac{\ln{\ln{L}}}{\ln{L}} \right)
\right\}
\; .
\label{ccFSSc}
\end{equation}
This was tested in \cite{KeIr97,Ja97}, 
but the lattice sizes available (up to $L=256$) were too small to  resolve the issue. 
(For a similar situation in the $q=4$ two-dimensional Potts model, see \cite{KiLa98}.)
The problem was revisited recently in \cite{Ha05} in which a very high precision numerical simulation 
achieved lattices as big as $L=2048$. 
Using the alternative ansatz 
\begin{equation}
\chi_L (0)  \sim L^{2-\eta_c} \left( C +  \ln{L} \right)^{-2r} 
\; ,
\label{ccFSSH}
\end{equation}
with $C$ an additional free parameter, this analysis gave $r=-0.056(7)$ provided suitably large
lattices are used (with $L \ge 256$). 
This value is compatible with the analytic prediction $r=-0.0625$.
In \cite{WeJa05}, a very careful study of the $F$ model, which is exactly solvable and has
a BKT-type phase transition, further elucidates the necessity  using sufficiently large 
lattices when numerically analysing models with highly subtle logarithmic corrections.

It is interesting to note that, with the exception of \cite{ToOk02}, all of the FSS-based
analyses of table~4 yield a negative value for $r$, in line with the BKT prediction.
It is rather surprising that,  in table~4, all of the thermal scaling analyses 
(as well as \cite{ToOk02}) yield positive values of $r$ -- far from the RG prediction
that $r = -0.0625$. This indicates that the FSS approach may generally be more powerful than the
thermal scaling one, although more extensive computational analyses  would
be required to  compare this approach to   \cite{Ha05}. 

Recently, Berche et al. have used a conformal map to rescale distances on the lattice
so that it is less sensitive to finite-size boundary effects and then to determine 
$\eta (\beta)$ at any temperature in the critical phase \cite{BeSh04,BeFa02,Be03}. 
In particular, their numerics 
yield accurate agreement with the analytic prediction $\eta_c = 1/4$ at the transition point.
It is interesting to note that this agreement was achieved very straightforwardly using 
moderate computational effort and
without recourse to enormous lattices.
Rather surprisingly, the evidence showed no noticable logarithmic corrections to the order-parameter 
density profile at the transition \cite{BeFa02,Be03} (see also \cite{HaRo98PaMe02}), 
but was clearly supportive of the existence
of such corrections in the correlation function \cite{BeSh04,Be03}.
However, the lattices used were not sufficiently large to unambiguously 
confirm the quantative behaviour of the logarithmic corrections.

\subsection{BKT vortex unbinding}

The second crucial aspect in the BKT scenario is the vortex unbinding mechanism. 
For the $XY$ model the energy of a single isolated 
charge-$n$ vortex on an $L\times L$ lattice is proportional to 
$n^2 \ln{L/a}$ where $a$ is the lattice spacing
(which never vanishes in a real physical system made up of atoms). 
The total energy of two vortices of charge $n$ and $-n$
centred at $x_1$ and $x_2$ is, however, proportional only to $n^2 \ln{|x_1-x_2|/a}$. At
low temperatures, then, vortices occur mostly in vortex-antivortex (dipole) pairs. 
Mutual cancellation of their individual ordering effects means that such 
a pair can only affect nearby spins and 
cannot significantly disorder the entire system. Topological long-range order exists in the system 
at low temperature.
As the temperature is raised, then, the  vortices proliferate and the distance between erstwhile
partners becomes so large that they are effectively free.
I.e., the BKT transition is one 
from a phase of dipoles to a plasma of vortices, which render the system disordered.

It was long  believed that adjusting the vortex-binding dynamics of the $XY$ model  may
disable the BKT transition mechanism, leading to a different one or the absence of a transition of 
any type \cite{SaWi88}.
With this in mind, the  step model is derived from the $XY$ model 
by replacing the  Hamiltonian (\ref{XY}) with 
\begin{equation}
 E = - \sum_{\langle i, j \rangle }{{\rm{sgn}}\left( \vec{s}_i\vec{s}_j \right) }
\;. 
\label{step}
\end{equation}
While the configuration space for this model is globally and continuously symmetric, as in the 
$XY$ case, its interaction function is discontinuous.
The energy associated with a single vortex for this system is 
expected to be independent of the lattice extent, leading to the expectation that
the disordered vortex-plasma phase may exist for all temperatures.
If this is the case, there was expected to be no vortex-driven phase transition in the model; 
if there is a phase transition, it was expected not to be of the BKT type.
Indeed, early studies supported this assertion (see \cite{KeIr97} for a review up to 1997).

In \cite{KeIr97,KeIr96}, strong and clear evidence was presented that a transition exists
in the model, and, that it belongs to the same universality class as the $XY$ model in two dimensions
(with even the corrections to scaling, insofar as they could be discerned, 
being the same as those for the $XY$ model).
Because of the dissimilar vortex energetics of the two models, this came as a surprise.
The question then arose as to how a transition mediated by vortices (in the $XY$ model)
can be insensitive to the energetics of such vortices.

In \cite{IoSh02}, the Mermin-Wagner theorem was extended to discontinuous
interaction functions including that of the step model (\ref{step}).
The issue of the phase transition there 
was again addressed in \cite{OlHo01}, where further evidence for the existence
of a BKT transition in the step model was  presented. 
In \cite{OlHo01}, the approach was to focus  on numerical analyses of the helicity modulus, 
which experiences a jump at the transition 
(for a similar approach to the $XY$ model see \cite{MiKi03}).
The main point of \cite{OlHo01} is that the cost in {\emph{free}} energy for 
a single vortex is proportional to $\ln{L/a}$ 
and that this is the feature which stabilizes the low-temperature (dipole) phase.
I.e., this inhibits proliferation of free vortices in the low-temperature phase.
It further explains the occurence of a transition in the
step model. 
The support for the 
existence of a BKT transition in the step model  given in \cite{OlHo01}
indicates that the BKT vortex scenario is more general than was heretofore realized.

\section{The diluted $XY$ model}
\label{dil}

A vibrant current area of research is the question of the role and consequences
of the presence of impurities in 
various systems, including the $XY$ model.
The occurrence of physical impurities renders any model more realistic, 
as such defects are often present in actual (and porous) systems. 
These physical impurities are modelled by removing (diluting) the sites or bonds
of the lattice.
Clearly, if the dilution is sufficiently strong  the percolation of spin-spin interactions
across the lattice is curtailed and it is effectively broken into finite disconnected sets.
This occurs at the percolation threshold.
In that circumstance, no phase transition can occur for any model, as a true phase transition
necessitates a thermodynamic limit. 
More moderate dilution is generally expected to lower the transition temperature
relative to its value for a pure (undiluted) system.

The special 
feature of the $XY$ model is the presence of vortices as the mechanism  mediating the transition.
It turns out that vortices are attracted to the physical impurities (vacant sites or bonds) 
and the vortex energy is reduced at such a vacancy. 
Therefore as the dilution is increased, more vortices can be formed and the consequent 
amount of vortex-induced disorder in the system is increased. 
This in turn may enhance the lowering of the critical temperature to such an extent that 
it vanishes even before the percolation threshold is reached. 

This  issue is addressed in \cite{Ro99HoDu02,LeCo03,BeFa03,WyPe05}.
According to the Harris criterion, disorder does not change the leading scaling behaviour of a model
if the critical exponent $\alpha$ associated with the  specific heat of the pure model 
is negative \cite{Ha74}. 
This is the case for the $XY$ model in two dimensions \cite{Ro99HoDu02}.
Thus the critical temperature in the diluted $d=2$ $XY$ model can be safely
(albeit approximately)
identified as being the location at which $\eta(\beta)=1/4$.
The vanishing of 
the critical temperature then gives  the critical vacancy density in that model.
In \cite{LeCo03}  the critical temperature was reported to vanish 
at site-vacancy density $\rho \approx 0.3$.
However, for $d=2$ regular lattices, the percolation threshold
(the density of vacancies required to disconnect the lattice)
 is $\rho \approx 0.41$.
Thus the work of \cite{LeCo03} suggests that, indeed, the critical temperature vanishes
before the percolation threshold is reached.
In a later study, however,  Berche et al.  suggested that the critical site-density 
is  closer to the percolation threshold \cite{BeFa03}. 
Support for the latter result also recently appeared in  \cite{WyPe05}.
These more recent studies -- which concern the {\emph{site}\/}-diluted version of the model --
indicate that the vortices do not, in fact, strongly enhance the lowering of the 
critical temperature in the $d=2$ $XY$ model.


This conclusion is further supported by a similar recent study \cite{SuOk05} for the 
{\emph{bond}\/}-diluted
$XY$ model, which favours the general 
scenario depicted in \cite{{BeFa03,WyPe05}} over that of  \cite{LeCo03}.

The above studies concern the value of the critical temperature in diluted 
two-dimensional $XY$ models.
The precise scaling behaviour of the thermodynamic functions at the phase transition is also of interest.
The Harris criterion concerns only the leading scaling behaviour 
(which does not alter in the $XY$ models considered here) and does not predict what 
effect dilution can have on the quantitative nature of the
exponents of the logarithmic corrections in the negative-$\alpha$ case.
This would be an interesting avenue for research in the future, and the two-dimensional
$XY$ model offers an ideal platform upon which to base such pursuits \cite{Ke05}.

\section{$O(n)$ and RP$^{n-1}$ models}
\label{Onnnn}

In this section, a selection of 
recent results in non-Abelian $O(n)$ models in two dimensions and in RP$^{n-1}$ models
in two and three dimensions are discussed, focusing on certain aspects that are still 
unresolved. 
For work on the $O(n)$ models in three (as well as two) dimensions,  the reader
is referred to the literature \cite{PV}.

\subsection{Non-Abelian $O(n)$ models in two dimensions}
\label{On}

It is widely believed that there are fundamental differences between models with 
Abelian and non-Abelian symmetry groups. 
The $O(2)$ symmetry group of the $XY$ model is Abelian
and all $O(n)$ groups with $n > 2$ are non-Abelian.
From the Mermin-Wagner theorem,  
any continuous symmetry of the $O(n)$ type cannot be broken in two dimensions \cite{MW}.
On this basis, there cannot be a transition to a phase with long-range order
in any $O(n)$ model with $n\ge 2$ there. 
(The $n=1$ case is discrete, and the corresponding Ising 
model possesses an ordered phase in $d=2$ \cite{Onsager}.)
As we have seen, in a two-dimensional theory, topological defects of dimension $m$ can exist 
if the $(1-m)^{\rm{th}}$ homotopy group, $\pi_{1-m}$, of the order-parameter space
(which for $O(n)$ models is the hypersphere $S^{n-1}$) is non-trivial. 
From table~\ref{tab1}, the only non-trivial homotopy group of the form $\pi_{1-m}(S^{n-1})$ 
is $\pi_{1}(S^{1}) \cong \Z$. 
This is the condition that gives rise to point defects (vortices) 
with integer charge in the  $n=2$ case (the $d=2$ $XY$ model).
The binding of these vortices at low temperature is the mechanism giving rise to the BKT phase transition 
\cite{B,KT}.

For $n>2$, therefore, conditions are not supportive of the existence of stable
topological defects of this type
in two dimensions  and
the majority belief is that there is no distinct low-temperature phase and consequently 
no positive-temperature phase transition in these models \cite{Po75}.
There is a vast literature on the $d=2$ $O(n)$ models with $n>2$ and the reader
is referred to  \cite{PV} for a review.
Furthermore, while perturbation theory predicts that these $n>2$  models are asymptotically free,
there is no rigorous proof to this effect.
(Asymptotic freedom means that the effective strength of interactions vanishes as the energy is increased.)
This notion has been  questioned and in \cite{Pa01,PaSe96,PaSe92}
evidence in support of the existence of phase transitions 
of the BKT type in these models has been given,
as well as heuristic explanations of why such transitions could occur
and a rigorous proof that this would be incompatible with asymptotic freedom.

Perturbative and high-temperature series expansions \cite{BuCo93}
as well as Monte Carlo calculations for $n\ge3$ \cite{AlBu99} have been performed, 
and these do not support the existence of such a BKT-like phase transition in these two-dimensional models.
Instead they are in agreement with perturbation theory and the asymptotic freedom scenario.
Nonetheless, the controversy has not been entirely resolved \cite{Se03} and
it is plausible that 
inclusion of logarithmic considerations in these considerations
could help in the search for a precise unambiguous resolution.

\subsection{Liquid crystals and RP$^{n-1}$ models}
\label{NLC}

The liquid-crystal state is a phase of matter which exists  distinct from (i.e., 
not a mixture of) the solid and liquid states. 
Unlike in a normal liquid, which is isotropic, in a liquid crystal
the properties are directional dependent. This is because the molecules of the liquid 
crystal are elongated in one direction in three-dimensional physical space. 
On the lattice, each molecule may be represented by a directed rigid rod. Such a direction
is without orientation, so that the system is unchanged by flipping the rod.
For the $n$-component model the Hamiltonian is
\begin{equation}
 E = - \sum_{\langle i,j \rangle}{\left(\vec{s}_i \vec{s}_{j}\right)^2}\,,
\label{lc}
\end{equation}
where $\vec{s}_i$ are $n$-component unit vectors and the quadratic form gives that
$-\vec{s}_i$  and $\vec{s}_i$ describe the same direction \cite{LL}.
The symmetry group for this $n$-vector model is ${\rm{RP}}^{n-1}$.
This is the space in $n$ dimensions formed by identifying antipodal points on an
$(n-1)$-sphere  $S^{n-1}$. Alternatively, restricting to one of the hemispheres
of $S^{n-1}$, the real projective space ${\rm{RP}}^{n-1}$ can be considered
as equivalent to the $(n-1)$-disk, $D^{n-1}$, with antipodes on the boundary $S^{n-2}$
identified (see example~4 of section~\ref{fund}).

For $n=2$, the real projective space ${\rm{RP}}^{1}$ is topologically equivalent to the circle $S^1$ in
$2$-space and the trigonometric
identity $2 \cos^2{\theta}= 1 + \cos{(2 \theta)}$ leads to the model (\ref{lc}) being equivalent to the 
$XY$ model.

In three dimensions, then, the condition for the existence of topological defects 
in the $n$-component model
is the non-triviality of
$\pi_{2-m}({\rm{RP}}^{n-1})$. 
Since ${\rm{RP}}^{n-1}$ is connected and $\pi_{0}({\rm{RP}}^{n-1})$
is trivial,  
there are no two-dimensional (wall) defects in a three-dimensional medium for any $n$.
In the liquid crystal case where $n=d=3$, table~\ref{tab2}, gives that 
$\pi_{2}({\rm{RP}}^2) \cong \Z$ so that point defects (monopoles,
called hedgehogs in this instance) may exist. (In addition to hedgehogs, if the system is finite
in extent, there may exist pointlike topological entities called boojums, which loosely speaking
are like half-hedgehogs -- see \cite{IsLa01} and references therein.)
Also from table~\ref{tab2}, since $\pi_{1}({\rm{RP}}^2) = \Z_2$, 
line defects (called disclinations) can  exist in the $n=d=3$ system. Indeed, 
this is reflected in the tendency of such molecules to align themselves into linear or
threadlike patterns. 
Such materials, which can cause polarization of light, are called nematic liquid crystals 
after the Greek prefix `nemato' meaning threadlike. Actually there are other types of liquid crystal 
states beyond the nematic one and the reader is refered to the (vast) literature
\cite{DeGo98}.

Similarly, in two dimensions, point defects (vortices) owe their existence to the non-triviality of 
$\pi_{1}({\rm{RP}}^{n-1})$, which is isomorphic to $\Z_2$. Even though only point defects can exist in the 
two-dimensional version, one often also generically refers to a nematic phase here too.

Perturbative RG analyses of the RP$^{n-1}$ model (which neglect the effects
of topological defects) predict no transitions in two dimensions 
\cite{Po75}
and a second-order one in three dimensions for $n\ge 3$ \cite{transin3D}. 
However, numerical and experimental 
research indicates that the three-dimensional RP$^{2}$ transition
is, in fact, weakly first-order \cite{LL,LaRo95,more2Dlc}. 
The three-dimensional RP$^{3}$ model also has a first-order transition \cite{Ro02}
and is closely linked to frustrated spin systems
which have been extensively studied (for a review of theoretical, numerical and experimental
work, see \cite{DeMo04}). 
The large $n$ limit of the $d=3$ RP$^{n-1}$ model also possesses a first-order transition \cite{KuZu89}.
It seems clear that topological
defects are responsible for 
the first-order phase transition in both the  three-dimensional RP$^{n-1}$ models
and their frustrated magnetic counterparts \cite{DeMo04}.

Since the RG approach fails in the three-dimensional 
model (where there is, in fact, 
a first-order transition), it is possible that it also fails in two dimensions 
and that this failure is due to the 
transitions being topologically driven \cite{KuZu92}.
Indeed, it has been shown that, in two dimensions, while perturbative RG predictions match well with
Monte Carlo measurements when the model has trivial homotopy, there is a clear
disagreement between the two approaches when topological defects are present \cite{Zu95}.

The two-dimensional RP$^{n-1}$ models were considered in \cite{KuZu92}, with $n=3$ and $n=40$. 
In the nematic $n=3$ case, evidence was presented for a transition  
described by a diverging correlation length and susceptibility but  a cusp (as opposed
to a divergence) in the 
specific heat was reported. Similar to the two-dimensional $XY$ case, 
both the correlation length and the susceptibility appeared to remain 
infinite below the critical temperature. 
Despite the scale of the study performed in \cite{KuZu92}, it was not possible to distinguish
between essential scaling and standard power-law scaling as the transition is approached from the
high-temperature phase.
The importance of the defects was, however,  clearly demonstrated,
in that they carry most of the energy and the transition appears to be again mediated by their unbinding
with increasing temperature. 
The $d=2$, $n=3$ case was also analysed in \cite{MoRo03}, where a second-order phase transition
was favoured.
In \cite{KuZu92}, a numerical analysis of the $n=40$ case was
also compared with analytical results for $n=\infty$. The latter has a topologically
mediated first-order transition for all $d \ge 2$ \cite{KuZu89}.
The possibility of (lattice-dependent) first-order transitions at large or infinite $N$
was discussed in \cite{late}.

The powerful conformal techniques used in \cite{BeSh04,BeFa02,Be03} were similarly employed in
\cite{BertrandRP2}, favouring a nematic/isotropic topologically-mediated transition
in the two-dimensional RP$^2$ model and a close similarity with the two-dimensional 
$XY$ model.
The effect of the suppression of the topological defects was explored in \cite{DuRo04}
where it was demonstrated that the apparent phase transition may be completely eliminated.
The suppression is achieved by the introduction of a chemical potential term associated with the
defects, making the formation of topological charges energetically expensive.

\noindent
\begin{table}[t]
\caption{Summary of the status of standard $O(n)$ and RP$^{n-1}$ models and some 
recent developments considered in this paper. Here $m$ is the defect dimension.
}
\label{taby}
\begin{center}
\begin{small}
\vspace{0.7cm}  
\noindent\begin{tabular}{|l|l|l|l|} 
    \hline     \hline
      & & &   \\
     Model  &$\!\!\!$ $d$  & $\!\!\!$ Homotopy group,$\!\!\!\!$  & Status  \\
            &               & $\!\!\!$ defect type (and $m$) $\!\!\!\!$  &  \\
    \hline
 & & & \\
  $XY$/${\rm{RP}}^1$/$O(2)$&   $2$ & $\Z$,    vortices ($m=0$)  &  BKT transition (see section~\ref{KT})    \\
  \& step model    &      &                     &          \\
 & & & \\
$O(n)$&  $2$ & No defects         &  No transition (majority opinion -- e.g.
                                                           \\
for $n \ge 3$  &       &   &  \cite{BuCo93,AlBu99,Po75,PV}, but see also  \cite{Pa01,PaSe96,PaSe92,Se03})\\
 & & &  \\
RP$^{n-1}$   &$2$& $\Z_2$, vortices $\forall~n$  & No transition \cite{Po75} (perturbation theory);  \\
for $n \ge 3$     &       & ($m=0$) & 1$^{\rm{st}}$-order transition \cite{KuZu89,late} ($n\rightarrow \infty$);  \\
  &        &                         & BKT or 2$^{\rm{nd}}$-order transition  \\
              &        &                          & for  $n=3$ \cite{KuZu92,MoRo03,BertrandRP2,DuRo04};     \\
              &        &                               &  No transition for $n=4$ \cite{CaAz01} \\
 & & &  \\
     $O(n)$   & $3$   & $n=2$: $\Z$, vortices         &  2$^{\rm{nd}}$-order transitions       \\
              &        &  ($m=1$);  $n=3$: $\Z$,          &  (see \cite{PV} and references therein)\\
              &        &  monopoles ($m=0$);           &                                         \\
              &        & no defects for $n\ge4$         &                                         \\
 & & & \\
RP$^{n-1}$     &$3$ & $n=3$ only: $\Z$,   &  2$^{\rm{nd}}$-order transition \cite{transin3D} (from\\
for $n \ge 3$ &    & monopoles  ($m=0$);              &   perturbation theory);    \\
 &   &  $n \ge 3$: $\Z_2$, vortices/             & 1$^{\rm{st}}$-order transition \cite{KuZu89} ($n\rightarrow \infty$);\\
              &       &  disclinations  ($m=1$)                         &  1$^{\rm{st}}$-order transition \cite{LL,LaRo95,more2Dlc} ($n=3$);\\
       &       &                                      & 1$^{\rm{st}}$-order transition \cite{Ro02,DeMo04} ($n=4$)\\
 & & &  \\
    \hline
    \hline 
  \end{tabular}
\end{small}
\end{center}
\end{table}
The suppression of the monopoles  in the three-dimensional 
Heisenberg model (which is also non-Abelian) also
leads to the disappearance of the transition there \cite{LaDa89HoJa94}. 
Similar work was applied to the nematic/isotropic transition of the three-dimensional 
RP$^{2}$ model in \cite{LaRo95}.
The first-order transition between the nematic and isotropic phases weakens as the 
disclination core energy is increased
and eventually the transition line splits into two continuous ones separating three distinct phases
with a new topologically-ordered isotropic phase between the nematic and conventionally
isotropic ones.

However, 
while the work of \cite{KuZu92,MoRo03,BertrandRP2,DuRo04} favours a second-order or 
BKT-type transition with divergent correlation
length in the $n=3$ case, and (see also \cite{WiEv95}), another study \cite{CaAz01}
of the RP$^{n-1}$ defects in a $d=2$ model of tops 
favours the absence of a true phase transition there (at least for $n=4$). 
Instead it is claimed that there is a crossover in the correlation length.
There it is argued that the defects disorder the system for all temperatures and the 
correlation length remains finite.

Therefore, the situation in the $d=2$ RP$^{n-1}$ models is still not satisfactorily clear 
and the precise nature of the phase transitions in these
models is  still under question.

In table~\ref{taby}, a summary of the status of the standard 
$O(n)$ and RP$^{n-1}$ models and the associated homotopy groups 
 is given, together with a  selection of recent papers.

\section{Highly nonlinear models}
\label{nsm}

Universality is the notion that the existence and type of phase transition 
in a model, and the critical exponents that describe it, depend only on the dimension,
the symmetries present and the range of interaction.
One can broaden the scope the models discussed herein, while maintaining these
three characteristics,  by altering the interaction to the form 
\begin{equation}
 E = - \sum_{\langle i,j \rangle}{P_k\left(\vec{s}_i \vec{s}_{j}\right)}\,,
\label{nonstandard}
\end{equation}
where $P_k(x)$ is the $k^{\rm{th}}$ Legendre polynomial.
Then $k=1$ and $k=2$ correspond to the $O(n)$ and $RP^{n-1}$ models appropriately.
The $k=4$ version has $P_4(x)\propto 35x^4-30x^2+3$
and has the same symmetries as the $k=2$ version but a higher degree of nonlinearity.

In $d=3$ dimensions, the inclusion of the $P_4$ interaction enhances the first-order transition
present in the $k=2$ model for $n=3$ (i.e., the RP$^{2}$ model) \cite{3dnl}.
Similarly, in two dimensions, the $n=3$, $P_4$ model was found to have a strong first-order transition 
\cite{PaRo03}.
In two dimensions, the claimed (but disputed)
continuous  transition in the 
$n=3$, $P_2$ model (i.e., the RP$^2$ model) and the first-order transition in the 
$n=3$, $P_4$ model can be eliminated by suppression of defects  \cite{DuRo04}.
Finally, in two dimensions, a BKT-like transition was recently observed in the $n=2$,
$P_4$ model in \cite{FaPa05}.
There, the crucial role of topology in such models is again exhibited and new questions are raised
regarding the detailed nature of that transition. These latest results demonstrate yet again the
importance and ubiquity of topologically mediated transitions in contemporary statistical mechanics.

\noindent
\begin{table}[th]
\caption{Summary of the status of some highly (i.e., sufficiently) nonlinear 
models and some recent developments considered in this paper.
Here $m$ is the dimension of the defects.
}
\label{tabz}
\begin{center}
\begin{small}
\vspace{0.7cm}  
\noindent\begin{tabular}{|l|l|l|l|} 
    \hline     \hline
      & & &   \\
     Model  &$\!\!\!$ $d$  & $\!\!\!$ Homotopy group,$\!\!\!\!$  & Status  \\
            &               & defect type (and $m$) &  \\
    \hline
 & & & \\
 $n=2$, $P_4$   & $2$  & $\Z$, vortices ($m=0$)   & BKT-like transition \cite{FaPa05} \\
 & & & \\
 $n=3$, $P_4$   & $2$  & $\Z_2$, vortices ($m=0$)  & $1^{\rm{st}}$-order transition \cite{DuRo04,PaRo03} \\
 & & & \\
 $n=3$, $P_4$   & $3$  & $\Z$, monopoles ($m=0$)$\!\!\!$  & $1^{\rm{st}}$-order transition \cite{3dnl} \\
                &       & $\Z_2$, vortices ($m=1$)  &                                              \\
 & & & \\
 Nonlinear $O(n)$  &  $2$ & $n=2$: $\Z$, vortices  ($m=0$);    &  1$^{\rm{st}}$-order transition for \\
                   &       &  no defects if $n\ge 3$                     & $n=2,3,\infty$ \cite{DoSh84,BlGu02,CaMo05}   \\
                   &       &  &  and generally \cite{vaSh02,vaSh05}  \\
 & & & \\
 Nonlinear RP$^{n-1}$ &  $2$ & $Z_2$, vortices ($m=0$)     &  1$^{\rm{st}}$-order transition \cite{vaSh05} \\
    &       &      &     \\
 Nonlinear $O(n)$ &  $3$ & $n=2$: $\Z$, vortices ($m=1$);         &  1$^{\rm{st}}$-order transition \cite{vaSh02} \\
                   &      & $n=3$: $\Z$, monopoles ($m=0$);   &  \\
                   &      & no defects for $n\ge 4$    &  \\
 & & &  \\
 Nonlinear RP$^{n-1}$ &  $3$ & $n=3$ only: $\Z$, monopoles ($m=0$);      &  1$^{\rm{st}}$-order transition \cite{vaSh05} \\
                   &      & $n\ge 3$: $\Z_2$, vortices ($m=1$)   &  \\
              &       &      &   \\
    \hline
    \hline 
  \end{tabular}
\end{small}
\end{center}
\label{tab40}
\end{table}
Recently another type of non-standard $n$-vector model has come into vogue.
This is given by a nonlinear
Hamiltonian of the type
\begin{equation}
 E = - \sum_{\langle i,j \rangle}{\left(1+\vec{s}_i \vec{s}_{j}\right)^p}\,.
\label{nonstandardd}
\end{equation}
If $p=1$, this recovers the standard $O(n)$ models.
There, for $d=2$, the $n=2$ version has the BKT transition discussed above, while 
for $n>2$ there is no positive-temperature transition.
In \cite{DoSh84}, \cite{BlGu02} and \cite{CaMo05} 
numerical evidence for the existence of first-order transitions in the $n=2$, $n=3$ 
and $n=\infty$ (spherical) models respectively 
in $d=2$ dimensions for sufficiently large $p$ was proffered.
A rigorous proof of the existence of first-order transitions in these
models was given by van~Enter and Shlosman \cite{vaSh02}
These transitions can occur in $d=2$ dimensions despite the implications for
zero magnetization coming from the Mermin-Wagner theorem \cite{MW}
and despite the high-$p$ models sharing the same 
dimension, symmetries range of interaction as the standard $p=1$ versions.

In three dimensions, where the Mermin-Wagner theorem does not apply, 
the standard $O(n)$ models, given by (\ref{nonstandardd}) with $p=1$,
have second-order transitions described by $n$-dependent critical exponents.
Again, the traditional notions of universality would imply that such behaviour 
is independent of $p$. However, sufficient nonlinearity (sufficiently large $p$)
can cause first-order transitions in these models
and the rigorous proof of \cite{vaSh02} extends to these $d=3$ cases too.

In \cite{vaSh05}, it is rigorously shown that various sufficiently nonlinear 
models of the RP$^{n-1}$ type also exhibit 
first-order transitions in $d\ge 2$  dimensions. Here, the Hamiltonian is of the form
\begin{equation}
 E = - \sum_{\langle i,j \rangle}{\left(1+(\vec{s}_i \vec{s}_{j})^2\right)^p}\,.
\label{nonstandard2}
\end{equation}
From these recent developments, it is clear that the notion of universality has to be extended
to include the degree of nonlinearity as a significant factor.
For an overview of recent developments in nonlinear models, see \cite{vaSh05b}
and table~\ref{tabz}.

\section{Conclusions}
\label{conc}

Topology plays an important role in condensed matter physics. In particular, 
homotopy theory facilitates the understanding and classification of the conditions 
which permit the existence of topological defects -- domain walls, vortices, monopoles and so on.  
After an expeditious introduction to the essentials of homotopy theory, a variety of 
topologically mediated phase transitions has been surveyed, starting with the 
BKT transition in the $XY$ model in two dimensions -- the paradigm for such studies.
After three decades of work, the detailed perturbative RG predictions of \cite{B,KT} have
been confirmed in that model through taking multiplicative logarithic corrections into account.
Furthermore, recent progress in confirming the vortex-binding mechanism as that
mediating the phase transition in the $XY$ model has been reviewed to complete an account of 
the current status of one of the most beautiful and remarkable models of theoretical physics.

Besides the $XY$ model, a number of other phase transitions have been
briefly examined, some of which are topologically-mediated. 
Some open problems have been highlighted, the resolution of which now stands at the
forefront of modern analytical and numerical investigations into statistical mechanics.

\section*{Acknowledgements}
The author wishes to thank Bertrand Berche, Arnaldo Donoso and Ricardo Paredes for 
their organisation of the Mochima Spring School. The remaining participants,
and in particular the students, are also thanked for making the School so dynamic
and interesting. The author further thanks Dominique Mouhanna and Aernout~van~Enter
for enlightening comments.


%
%
  \label{last@page}
  \end{document}